\documentstyle[11pt]{article}
\input epsf
\newcommand{\be}{\begin{eqnarray}}
\newcommand{\ee}{\end{eqnarray}}
\newcommand{\la}[1]{\label{#1}}

 \def\Tr{\mbox{Tr}}

\def\Dirac#1{\hat{#1}}
\def\ns{\Dirac n}
\def\dd{\Dirac \partial}
  \def\beq{\begin{equation}}
  \def\eeq{\end{equation}}
\begin{document}
%
%
\rightline{RUB-TPII-19/98}
\vspace{.3cm}
\begin{center}
{\Large \bf Helicity  skewed quark distributions of the nucleon \\
and chiral symmetry}
\\
\vspace{1.4cm}
{\large
M.\ Penttinen$^a$,
M.V.\ Polyakov$^{a,b}$ and K. Goeke$^{a}$}
\\[0.5 cm]
$^a${\em Institut f\"ur Theoretische Physik II,
Ruhr--Universit\"at Bochum, \\ D--44780 Bochum, Germany} \\[0.2cm]
$^b${\em Petersburg Nuclear Physics
Institute, 188350 Gatchina, Russia}
\end{center}
\vspace{.6cm}
\begin{abstract}
\noindent
We compute the helicity skewed quark distributions $\widetilde{H}$
and $\widetilde{E}$ in the chiral quark-soliton model
of the nucleon. This model
emphasizes correctly the role of spontaneously
broken chiral symmetry in structure of nucleon.
It is based on the large-$N_c$ picture of the
nucleon as a soliton of the effective chiral lagrangian and allows to
calculate the leading twist quark-- and antiquark distributions at a low
normalization point.

We discuss the role of chiral symmetry  in the
helicity skewed quark distributions $\widetilde{H}$ and $\widetilde{E}$.
We show that generalization of soft pion theorems, based on
chiral Ward identities, leads in the region of $-\xi<x<\xi $
to the pion pole contribution
to $\widetilde{E}$  which dominates
at small momentum transfer.
\end{abstract}

\newpage

\renewcommand{\theequation}{\arabic{section}.\arabic{equation}}

\setcounter{equation}{0}

\section{Introduction}

Recently, a new type of parton distributions
\cite{Leipzig,Ji1,Rad,CFS} has attracted
considerable interest, the so-called
skewed parton distributions
(SPD's), which are generalizations simultaneously of the usual parton
distributions, distribution amplitudes
and of the elastic nucleon form factors (for review see \cite{JiReview}).
Taking the $n$--th
moment of the SPD's one obtains the form factors ({\em i.e.},
non-forward matrix elements) of the spin--$n$, twist--two quark and
gluon operators.  On the other hand, in the forward limit the SPD's
reduce to the usual quark, antiquark and gluon distributions. In other
words, the SPD's interpolate between the traditional inclusive (parton
distributions) and exclusive (form factors) characteristics of the
nucleon and thus provide us with a considerable new amount of
information on nucleon structure.

The SPD's are not accessible in standard inclusive measurements. They
can, however, be measured in deeply--virtual Compton scattering (DVCS)
and in hard exclusive leptoproduction of mesons. The very possibility
to probe SPD's in these reactions is due to QCD factorization theorem
of Ref.~\cite{CFS}.
Feasibility of experimental measurements of SPD's in hard
exclusive reactions
is currently being studied \cite{Munich,Freu,MV,MPR98}.
A quantitative description of these classes of processes requires
non-perturbative information in the form of the SPD's at some initial
normalization point.
Although the skewed parton distributions can be reduced
in certain limiting cases
to already known
quantities (parton distributions, form factors),  even their
qualitative behaviour  is unknown to large extent.  That is why model
calculations of these quantities are of big importance.  There were
already model calculations of SPD's: in the bag model \cite{Ji3} and in
the chiral quark-soliton model \cite{my}. In the latter calculation a
drastic variation of flavour singlet  $H(x,\xi,t)$ at $x$ near
$x=|\xi|$ was observed. Such behaviour is related to the fact that the
SPD's in the region $-\xi<x<\xi$ have properties of distribution
amplitudes. This feature, being very important for the understanding
of SPD's, requires a field theoretic description of nucleon's
constituents and that is the reason why it can not by reproduced
in the bag model.

Our aim now is to compute helicity skewed quark distributions
of the nucleon using the methods of ref.~\cite{my}.
We shall see that generalization of
low energy theorems  requires that the skewed distribution
$\widetilde E$ develops a pion pole\footnote{
The contribution of the pion pole to $\widetilde E$
was discussed at qualitaive level in ref.~\cite{NDCEBAF}}
at $\Delta^2=m_\pi^2$ of the form:

\be
\lim_{\Delta^2\to m_\pi^2}
\widetilde E^{(3)}_\pi(x,\xi,\Delta^2)=
-\frac{4 g_A M_N^2}{|\xi|(\, \Delta^2-m_\pi^2 \, )}\,
\theta \Big( |x|<|\xi| \Big)\
\Phi_\pi\Biggl(\frac{x}{\xi}\Biggr),
\ee
where $\Phi_\pi(z)$ is distribution amplitude of the pion.
In refs.~\cite{MPR98,FPPS} it was shown that this contribution
to $\widetilde E$ leads to considerable enhancement of the
amplitude of hard exclusive production of charged pions
and to large azimuthal spin asymmetry in
exclusive $\pi^\pm$ production \cite{FPPS}.

We shall see that in the chiral quark-soliton model the
pion pole contribution is related to the large distance asymptotic
of the pion mean-field, which is controlled by PCAC.

\section{Definition of skewed helicity quark distributions}
\label{section_QCD_def}

In QCD the helicity skewed quark distributions are defined through
non-diagonal matrix elements of product of quark fields at light--cone
separation. Here and in the following, we shall use the notations
of ref.~\cite{JiReview}

\be
\nonumber
\int \frac{d\lambda }{2\pi }e^{i\lambda x}
\langle P^{\prime }|\bar \psi_f
(-\lambda n/2){\Dirac n} \gamma_5
 \psi_f (\lambda n/2)|P\rangle &=& \widetilde{H}_f(x,\xi ,\Delta^2) \;
\bar U(P^{\prime }) \; \ns \gamma_5 \; U(P) \\
&+& \frac 1{2M_N} \; \widetilde{E}_f(x,\xi ,\Delta^2) \;
\bar U(P^{\prime }) \; (n\cdot  \Delta) \gamma_5 U(P) .
\nonumber \\
\label{E-H-QCD-2}
\ee
Here $n_\mu$ is a light-cone vector,
\be
n^2 &=& 0, \hspace{1.5cm} n\cdot (P + P') =2\ n\cdot\bar P
\; = \;\; 2,
\label{n-normalization}
\ee
$\Delta$ is the four--momentum transfer,
\begin{equation}
\Delta = P^{\prime }-P , \label{Delta-def}
\end{equation}
$M_N$ denotes the nucleon mass, and $U(P)$ is a standard Dirac
spinor. The skewed
quark distributions, $\widetilde{H}(x,\xi ,\Delta ^2)$
and $\widetilde{E}(x,\xi ,\Delta ^2)$,
are regarded as functions of the variable
$x$, the square of the four--momentum transfer, $\Delta^2=t$, and its
longitudinal component

\begin{equation} \xi =-\frac 12 \, (n\cdot \Delta ) .
\label{xi-def} \end{equation}

In the forward case, $P = P'$, both $\Delta$ and $\xi$ are zero, and
the second term on the r.h.s.\ of eq.(\ref{E-H-QCD-2}) disappears.
In this limit the
function $\widetilde{H}$ becomes the usual polarised parton distribution
function,
\be
\widetilde H_f \Big( x, \xi = 0, \Delta^2= 0 \Big) &=&
\left\{
\begin{array}{cr}
\Delta q_f(x) ,& \hspace{.5cm} x \; > \; 0\,, \\
\Delta \bar q_f(-x)
,& \hspace{.5cm} x \; < \; 0 \,.
\end{array}
\right.
\label{forward_limit}
\ee
On the other hand, taking the first moment of eq.(\ref{E-H-QCD-2})
one reduces the operator on the l.h.s. to the local
axial vector current.
The dependence of $\widetilde{H}$ and $\widetilde{E}$ on $\xi$
disappears, and the functions
reduce to the usual axial form factors of the nucleon,

\be
\int_{-1}^1 dx \;
\widetilde{H} (x, \xi , \Delta^2) &=& G_A (\Delta^2 ), \\
\int_{-1}^1 dx \;
\widetilde{E}(x, \xi , \Delta^2 ) &=& G_P (\Delta^2 ).
\label{srE}
\ee

Taking higher moments of the distribution functions one obtains the form
factors of the twist--2, spin--$n$ operators.

\section{Chiral quark--soliton model of the nucleon}
\label{section_model}

Recently a new approach to the calculation of quark distribution functions
of the nucleon
has been developed \cite{DPPPW} in the framework of the chiral
quark-soliton model of the nucleon \cite{DPP}. In present paper we apply this
approach to the calculation of skewed quark distributions.
It is essentially based on the
$1/N_c$ expansion.  Although in reality the number of colours $N_c=3$,
the academic limit of large $N_c$ is known to be a useful guideline. At
large $N_c$ the nucleon is heavy and can be viewed as a classical
soliton of the pion field \cite{Witten,ANW}.  In this paper we work
with the effective chiral action given by the functional integral over
quarks in the background pion field \cite{DE,DSW,DP}:

\begin{eqnarray}
\exp \Big( iS_{{\rm eff}}[\pi (x)]\Big) =\int D\psi D\bar \psi \;\exp
\left( i\int d^4x\,\bar \psi (i\dd -MU^{\gamma _5})\psi \right) ,
\nonumber
\end{eqnarray}
\begin{equation}
U\;=\;\exp \Big( i\pi ^a(x)\tau ^a\Big) ,\hspace{1cm}U^{\gamma
_5}\;=\;\exp \Big( i\pi ^a(x)\tau ^a\gamma _5\Big) \;=\;\frac{1+\gamma _5}
2U+\frac{1-\gamma _5}2U^{\dagger }.  \label{FI}
\end{equation}
Here $\psi $ is the quark field, $M$ is
the effective quark mass, which is
due to the spontaneous breakdown of chiral symmetry (generally
speaking, it is momentum dependent), and $U$ is the $SU(2)$ chiral pion
field. The effective chiral action given by (\ref{FI}) is known
to contain automatically the Wess--Zumino term and the four-derivative
Gasser--Leutwyler terms, with correct coefficients.
The equation (\ref{FI}) has been
derived from the instanton model of the QCD vacuum \cite{DP,DP1}, which
provides a natural mechanism of chiral symmetry breaking and enables one to
express the dynamical mass $M$ and the ultraviolet cutoff intrinsic in
(\ref{FI}) through the $\Lambda _{QCD}$ parameter. The ultraviolet
regularization of the effective theory is provided by the specific momentum
dependence of the mass, $M(p^2)$, which drops to zero for momenta of
order of the inverse instanton size in the instanton vacuum,
$1/\rho \sim 600\, {\rm MeV}$. For simplicity we shall neglect this momentum
dependence in the general discussion; it will be taken into account again
in the theoretical analysis and in the numerical estimates later.

An immediate application of the effective chiral theory (\ref{FI}) is the
quark-soliton model of baryons of ref.~\cite{DPP}, which is in the spirit of
the earlier works~\cite{KaRiSo,BiBa}. According to this model nucleons can
be viewed as $N_c$ ``valence" quarks bound by a self-consistent pion
field (the ``soliton") whose energy coincides with the aggregate
energy of the quarks of the negative-energy Dirac continuum. Similarly to
the Skyrme model large $N_c$ is needed as a parameter to justify the use of
the mean-field approximation; however, the $1/N_c$--corrections can be
--- and, in some cases, have been --- computed \cite{Review}.

Let us remind the reader how the nucleon is described in the effective
low-energy theory (\ref{FI}). Integrating out the quarks in (\ref{FI}) one
finds the effective chiral action,
\begin{equation}
S_{{\rm eff}}[\pi ^a(x)]=-N_c\,\mbox{Sp}\log D(U)\,,\hspace{1cm}
D(U)\;\;=\;\;i\partial _0-H(U),  \label{SeffU}
\end{equation}
where $H(U)$ is the one-particle Dirac Hamiltonian,
\begin{equation}
H(U)=-i\gamma ^0\gamma ^k\partial _k+M\gamma ^0U^{\gamma _5}\,,  \label{hU}
\end{equation}
and $\mbox{Sp}\ldots $ denotes the functional trace.
For a given time-independent pion field \mbox{$U=\exp(i\pi^a({\bf x})\tau^a)$} one
can determine the spectrum of the Dirac Hamiltonian,
\begin{equation}
H\Phi_n = E_n \Phi_n.  \label{Dirac-equation}
\end{equation}
It contains the upper and lower Dirac continua (distorted by the presence of
the external pion field), and, in principle, also discrete bound-state
level(s), if the pion field is strong enough. If the pion field has unity
winding number, there is exactly one bound-state level which travels all the
way from the upper to the lower Dirac continuum as one increases the spatial
size of the pion field from zero to infinity \cite{DPP}. We denote the
energy of the discrete level as $E_{{\rm lev}},\;\;-M\leq E_{{\rm lev}}\leq
M $. One has to occupy this level to get a non-zero baryon number state.
Since the pion field is colour blind, one can put $N_c$ quarks on that level
in the antisymmetric state in colour.

The limit of large $N_c$ allows us to use the mean-field approximation to
find the nucleon mass. To get the nucleon mass one has to add
$N_cE_{{\rm lev}}$ and the energy of the pion field.
Since the effective chiral lagrangian
is given by the determinant (\ref{SeffU}) the energy of the pion field
coincides exactly with the aggregate energy of the lower Dirac continuum,
the free continuum subtracted. The self-consistent pion field is thus found
from the minimisation of the functional \cite{DPP}

\begin{equation}
M_N = \min_U \; N_c\left\{E_{{\rm lev}}[U] \; + \;
\sum_{E_n<0} \Big( E_n[U]-E_n^{(0)} \Big) \right\}.  \label{nm}
\end{equation}
{From} symmetry considerations one looks for the minimum in a hedgehog ansatz:
\begin{equation}
U_c({\bf x}) \; = \; \exp\Big( i\pi^a({\bf x})\tau^a \Big) \; = \;
\exp\Big( i n^a \tau^a P(r) \Big), \hspace{1cm} r \; = \; |{\bf x}|,
\hspace{1cm} {\bf n} \; = \; \frac{{\bf x}}{r} ,  \label{hedge}
\end{equation}
where $P(r)$ is called the profile of the soliton.

The minimum of the energy (\ref{nm}) is degenerate with respect to
translations of the soliton in space and to rotations of the soliton field
in ordinary and isospin space. For the hedgehog field (\ref{hedge}) the two
rotations are equivalent.
The projection on a nucleon state with
given spin ($S_3$) and isospin ($T_3$) components
is obtained by integrating over all spin-isospin rotations,
$R$ \cite{ANW,DPP},
\be
\langle S=T,S_3,T_3|\ldots| S=T,S_3,T_3\rangle &=& \int
dR\;\phi^{\ast\;S=T}_{S_3T_3}(R) \; \ldots \; \phi^{S=T}_{S_3T_3}(R)\,.
\label{spisosp}
\ee
Here $\phi^{S=T}_{S_3T_3}(R)$ is the rotational wave function of the nucleon
given by the Wigner finite-rotation matrix \cite{ANW,DPP}:
\be
\phi _{S_3T_3}^{S=T}(R) &=& (-1)^{T+T_3} \sqrt{2S+1}\, D_{-T_3,S_3}^{\,S=T}(R).
\label{Wigner}
\ee
Analogously, the projection on a nucleon state with given momentum
${\bf P}$ is obtained by integrating over all shifts, ${\bf X}$, of the
soliton,
\be
\langle {\bf P^\prime}|\ldots|\ {\bf P}\rangle
&=& \int d^3{\bf X}\;e^{i \left( {\bf P^\prime-P} \right) \cdot{\bf X}}\; \ldots
\label{totmom}
\ee

\section{Skewed quark distributions in the chi\-ral
quark--so\-li\-ton model}
\label{section_SQD_in_model}

We now turn to the calculation of the skewed quark distributions in
the chiral quark--soliton model. This description of the nucleon
is based on the $1/N_c$-expansion. At large $N_c$ the nucleon is
heavy --- its mass is $O(N_c)$.
For the large-$N_c$ nucleon
eq.~(\ref{E-H-QCD-2}) simplifies as follows:
\be
\nonumber
\int \frac{d\lambda }{2\pi }e^{i\lambda x}
\langle P^{\prime },S_3^{\prime}|\bar \psi_f (-\lambda n/2){\Dirac n} \gamma_5
\psi_f (\lambda n/2)|P,S_3\rangle &  & \\
\,=\, \frac{\tau^3_{ff}}{2} \Bigg\lbrace \, 2\delta^{3i}
\widetilde{H}_f(x,\xi , t)
& - &\frac{\Delta^3 \Delta^i}{2 M_N^2} \, \widetilde{E}_F(x,\xi,t) \,
\Bigg\rbrace \,  \sigma^i_{S_3^{\prime }S_3} \label{E-H-QCD-3} \, ,
\ee
where $S_3, S_3'$ denote the projections of the nucleon spin.
{}From this expression
we immediately see that in the leading order of
the $1/N_c$-expansion only the flavour isovector part of
$$\widetilde{H}^{(3)}(x,\xi ,\Delta ^2)=
\widetilde{H}_u(x,\xi ,\Delta ^2)-
\widetilde{H}_d(x,\xi ,\Delta ^2)$$
and
$$\widetilde{E}^{(3)}(x,\xi ,\Delta ^2)=
\widetilde{E}_u(x,\xi ,\Delta ^2)-
\widetilde{E}_d(x,\xi ,\Delta ^2)$$
are non-zero.
The isosinglet part of
$\widetilde{H}(x,\xi ,\Delta ^2)$ and the isosinglet part
of $\widetilde{E}(x,\xi ,\Delta ^2)$ appear only in the next--to--leading
order of the $1/N_c$--expansion, {\em i.e.}, after taking into account
the finite angular velocity of the soliton rotation.

Before computing the skewed
quark distribution functions we must determine the
parametric order  in $1/N_c$ of the kinematical variables involved.
Generally, when describing parton distributions in the large--$N_c$ limit,
one has
$x \sim 1/N_c$, since the nucleon momentum is distributed
among $N_c$ quarks. Furthermore, as in the calculation of nucleon form
factors we
consider momentum transfers to be of order
$t \sim N_c^0$; hence, in particular,
$\xi \sim 1/N_c$, so that $\xi$ is of the same parametric order as
$x$.

Technically
the calculation of the skewed parton distributions proceeds in much
the same way as that of the usual parton distributions \cite{DPPPW,
my}.
Using the formalism developed in \cite{DPPPW,my} we obtain:
\begin{eqnarray}
\nonumber
\widetilde{H}^{(3)}(x, \xi, \Delta^2) \, =
&-&\frac{N_c M_N}{6\pi \mbox{\boldmath $\Delta$}_\perp^2}
\int dz^0 \int d^3{\bf X} \,
\exp \Big( i\mbox{{\boldmath $\Delta$}} \cdot {\bf X} \Big)
\sum\limits_{{\scriptstyle}\atop {\scriptstyle {\rm occup.}}}
\exp \Big( iz^0\Big[ (x+\xi)M_N-E_n \Big] \Big) \;
\nonumber \\
&\,& \Phi_n^{\dagger }({\bf X})
\Big(\mbox{\boldmath $\Delta$}_\perp^2 \tau^3+2\xi M_N
\mbox{{\boldmath $\Delta$}}_{\perp} \cdot \mbox{\boldmath $\tau$}_\perp \Big)
\Big( 1+\gamma^0 \gamma^3 \Big) \gamma_5 \, \Phi_n \, ({\bf X}-z^0{\bf e}_3),
\label{H-singlet-general}
\end{eqnarray}

\begin{eqnarray}
\nonumber
\widetilde{E}^{(3)} (x,\xi ,\Delta ^2) \, =
&-& \frac{N_cM_N^2}{3\pi\xi\mbox{\boldmath $\Delta$}_\perp^2}
\int dz^0 \int d^3{\bf X} \,
\exp \Big( i\mbox{{\boldmath $\Delta$}} \cdot {\bf X} \Big)
\sum\limits_{{\scriptstyle}\atop {\scriptstyle {\rm occup.}}}
\exp \Big( iz^0 \Big[(x+\xi)M_N-E_n \Big] \Big)  \\
&\, &\Phi_n^{\dagger }({\bf X})
\Big( \mbox{{\boldmath $\Delta_\perp$}}\cdot \mbox{{\boldmath $\tau_\perp$ }} \Big)
\Big( 1+\gamma ^0\gamma ^3 \Big)\gamma_5 \, \Phi_n \,({\bf X}-z^0{\bf e}_3).
\label{E-nonsinglet-general}
\end{eqnarray}

Before going ahead with the evaluation of the
expressions eqs.(\ref{H-singlet-general}, \ref{E-nonsinglet-general})
we would like to demonstrate that the two limiting cases
of the skewed distributions
--- usual parton distributions and elastic form factors --- are
correctly reproduced within the chiral quark--soliton model.
Taking in eq.(\ref{H-singlet-general}) the forward limit,
$\Delta\rightarrow 0$, one recovers the formula for the usual
polarized (anti--) quark distributions in our model which was
obtained in ref.~\cite{DPPPW}. Thus the forward limit,
eq.~(\ref{forward_limit}), is reproduced. On the other hand, integrating
eq.~(\ref{H-singlet-general}) over $-1\le
x\le 1$ one obtains (up to corrections parametrically small in $1/N_c$)
the expressions for the axial form factors of the nucleon
derived in ref.~\cite{mojdipl}:
\be
\nonumber
\int_{-1}^1 dx\ \widetilde H^{(3)}(x,\xi ,\Delta ^2) &=&
-\frac{N_c}{3}
\int d^3{\bf X} \, \exp \Big( i\mbox{{\boldmath $\Delta$}}\cdot {\bf X} \Big)
\sum\limits_{{\scriptstyle}\atop {\scriptstyle {\rm occup.}}}
\Phi_n^{\dagger }({\bf X})
\, \tau^3 \gamma^0\gamma^3\gamma_5 \, \Phi_n \,({\bf X})
\nonumber \\
&=& G_A^{\rm (T=1)}(\Delta^2)\, .
\label{H-singlet-sum-rule}
\ee
Actually experimental $G_A^{\rm (T=1)}(\Delta^2)$
is very well reproduced in the chiral quark-soliton model up to
momenta of order $\Delta^2\sim 1$~GeV$^2$ \cite{Review}.

Now if one integrates
eq.~(\ref{E-nonsinglet-general}) over $-1\le
x\le 1$ one obtains (up to corrections parametrically small in $1/N_c$)
the following expression:

\begin{equation}
\int_{-1}^1 dx \, \widetilde E^{(3)}(x,\xi ,\Delta ^2)=
-\frac{2 N_c M_N}{3\xi \mbox{\boldmath $\Delta$}_\perp^2}
\int d^3{\bf X}\exp \Big( i\mbox{{\boldmath $\Delta$}}\cdot {\bf X}\Big)
\sum\limits_{{\scriptstyle}\atop {\scriptstyle {\rm occup.}}}
\Phi_n^{\dagger }({\bf X}) \gamma^0 \gamma^3
\Big( \mbox{{\boldmath $\Delta_\perp$}} \cdot
\mbox{{\boldmath $\tau_\perp $}} \Big)
\gamma_5 \Phi _n({\bf X}).  \label{E-nonsinglet-sum-rule}
\end{equation}
Using the ``hedgehog" symmetry of the pion mean-field one can
easily show that the expression (\ref{E-nonsinglet-sum-rule})
is a function of only $\Delta^2$ and coincides with expression for
pseudoscalar nucleon form factor in the chiral quark soliton model,
see e.g. ref.~\cite{Review}.

Eqs.~(\ref{H-singlet-general}, \ref{E-nonsinglet-general}) express
the SPD's as a sum over quark single--particle levels in the
soliton field. This sum runs over {\it all} occupied levels, including
both the discrete bound--state level and the negative Dirac continuum.
We remind the reader that in the case of usual parton distributions
it was demonstrated that in order to ensure the positivity of the antiquark
distributions it is essential to take into account the contributions
of {\it all} occupied levels of the Dirac Hamiltonian \cite{DPPPW}.
We shall see below that also in the
case of skewed quark distributions
the contribution of the Dirac continuum drastically
changes the shape of the distribution function.
That is especially important
to reproduce the pion pole contribution to the spin-flip
SPD $\widetilde E^{(3)}$ required by chiral Ward identities.

The contribution of the discrete bound--state level to
eqs.~(\ref{H-singlet-general}, \ref{E-nonsinglet-general})
can be computed using the expressions
given in the Appendix. The result is shown in Figs.~1 for the forward case
and Figs.~2
for a non-zero momentum transfer. Being taken by itself this
contribution resembles qualitatively
the shape of SPD's $\widetilde H$
and $\widetilde E$ obtained in
the bag model \cite{Ji3}.

To calculate the contribution of the Dirac continuum to
eqs.~(\ref{H-singlet-general}, \ref{E-nonsinglet-general})
we resort to an appro\-xima\-tion
which proved to be very successful in the computation of usual parton
distributions, the so--called interpolation formula \cite{DPPPW}.
One first expresses the continuum contribution as a functional trace
involving the quark propagator in the background pion field. The
quark propagator can then be expanded in powers of the formal parameter
$\partial U/(-\partial^2+M^2)$, which becomes small
in three limiting cases: {\em i}) low momenta, $|\partial U|\ll M$,
{\em ii}) high momenta, $|\partial U|\gg M$, {\em iii}) any momenta but
small pion fields, $|\log U| \ll 1$. One may therefore expect that this
approximation has good accuracy also in the general case. As was shown in
refs.~\cite{DPPPW} for usual parton distributions this approximation
preserves the positivity of the antiquark distributions and all sum rules;
moreover, it gives results very close to those obtained by exact numerical
diagonalisation of the Dirac Hamiltonian and summation over the
negative--energy levels.

Simple generalization of the technique developed in
\cite{DPPPW,my} allows us to express the Dirac continuum contribution
to the SPD's directly in terms of pion mean-field
(\ref{hedge}):
\be
\nonumber
\widetilde H^{(3)}(x,\xi ,\Delta ^2)_{\rm cont} &=&
-\frac{ M_N N_c}{3\ \mbox{\boldmath $\Delta$}_\perp^2}
\; \mbox{Im} \, \int \frac{d^3k}{(2\pi )^3}\int
\frac{d^4p}{(2\pi )^4} \;
\delta \left[ \Bigl(x- \xi \Bigr) M_N-v\cdot p\right] \\
\label{H-1-sym-res-mp}
&\times &
\frac{M_1^{1/2}}{\Big( p_1^2-M_1^2+i0 \Big)} \,
\frac{M_2^{3/2}}{\Big( p_2^2- M_2^2+i0 \Big)}
\Big( k\cdot v+\frac 12 \Delta\cdot v \Big) \\
\nonumber
&\times& \Tr_{\rm fl.} \Biggl[ \Big( \mbox{\boldmath $\Delta$}_\perp^2\tau^3+2\xi M_N
\mbox{{\boldmath $\Delta_{\perp} $}} \cdot \mbox{{\boldmath $\tau_\perp$}}\Big)
\widetilde{U} \Big({\bf k-}\mbox{{\boldmath $\frac\Delta 2$}}\Big)
\Big[\widetilde{U} \Big( {\bf k+}\mbox{{\boldmath $\frac\Delta 2$}}
\Big)\Big]^{+} \Biggr] \\
\nonumber
&+&\Biggl( \xi \to -\xi, \;\; \mbox{{\boldmath $\Delta$}}
\to -\mbox{{\boldmath $\Delta$}}
\Biggr),
\ee

\be
\nonumber
\widetilde E^{(3)}(x,\xi ,\Delta ^2)_{\rm cont}
&=& -\frac{2 M_N^2 N_c}{3\ \xi\ \mbox{\boldmath $\Delta$}_\perp^2}
 \; \mbox{Im} \, \int \frac{d^3k}{(2\pi )^3}\int \frac{d^4p}{(2\pi )^4} \;
\delta \left[ \Bigl(x- \xi \Bigr) M_N-v\cdot p\right] \\
&\times &
\label{E-1-sym-res-mp}
\frac{M_1^{1/2}}{\Big( p_1^2-M_1^2+i0 \Big)}\,
\frac{M_2^{3/2}}{\Big( p_2^2-M_2^2+i0 \Big)}
\, \Big( k\cdot v+\frac 12 \Delta\cdot v \Big)\\
\nonumber
&\times & \Tr_{\rm fl.} \Biggl[ \Big( \mbox{{\boldmath $\Delta_{\perp} $}} \cdot \mbox{{\boldmath $\tau_{\perp} $}} \Big)
\widetilde{U}\Big( {\bf k-}\mbox{{\boldmath $\frac\Delta 2$}} \Big)
\Big[\widetilde{U} \Big({\bf k+}\mbox{{\boldmath $\frac\Delta 2$}}\Big) \Big]^{+} \Biggr]+\Biggl( \xi \to -\xi, \;\; \mbox{{\boldmath $\Delta$}}
\to -\mbox{{\boldmath $\Delta$}}
\Biggr),
\ee
where $v=(1,0,0,-1)$ is a light cone vector and
the Fourier transform of the soliton field is defined as

\begin{equation}
\widetilde{U}({\bf k} ) \, \equiv \,
\int d^3{\bf x\,} e^{-i {\bf k} \cdot {\bf x} \,}
U({\bf x}) \, .
\label{FT}
\nonumber
\end{equation}
Also we introduced short notation $p_1=p+\Delta$, $p_2=p-k-\Delta/2$,
$M_1=M(p_1^2)$ and $M_2=M(p_2^2)$.
In eqs.~(\ref{H-1-sym-res-mp}, \ref{E-1-sym-res-mp}) the momentum
dependence of the constituent quark mass, $M(p^2)$ cuts the
loop momentum $p$ and thus regularizes the UV divergence.
Let us note that expressions (\ref{H-1-sym-res-mp}, \ref{E-1-sym-res-mp})
are explicitly symmetric under transformation \mbox{$\xi\to-\xi$} what
follows from charge conjugation symmetry \cite{Munich,JiReview}.

\subsection{Pion pole contribution to the
skewed quark distribution $\widetilde E$}

Before presenting the numerical results for the SPD's
$\widetilde H$ and $\widetilde E$ let us discuss specific
contribution to these SPD's originating from the long range pion tail
of the pion mean-field.
The behaviour of the mean pion field at large distances is governed
by linearized equations of motion and PCAC:

\be
\lim_{|\vec x|\to \infty }U(x) \, =\, 1+ \frac{3 g_A}{8\pi f_\pi^2 |\vec x|^2}
\Big( 1+m_\pi |\vec x| \Big) \,
\frac{i x^a\tau^a}{|\vec x|}
\exp(-m_\pi |\vec x|)\; ,
\ee
where $g_A\approx 1.25$ is the axial charge of the nucleon,
$f_\pi\approx 93$~MeV is the pion decay constant.
This asymptotic implies that the Fourier transform
(\ref{FT}) has the following small momentum asymptotic:

\be
\widetilde U(\vec k) \, \sim \, (2\pi)^3 \delta(\vec k)+
\frac{3 g_A}{2 f_\pi^2}\frac{(\vec k\cdot \vec \tau)}{\vec k^2+m_\pi^2
}\, .  \label{Ytail} \ee
It is useful to split the Fourier transform
of the pion mean-field into two pieces:

\be
\widetilde U(\vec k) \, =\, \widetilde U(\vec k)_{{\rm smooth}}+
(2\pi)^3 \delta(\vec k)\, ,
\label{twopieces}
\ee
where
\be
\widetilde U(\vec k)_{{\rm smooth}} \, =\,
\int d^3{\bf x\,}e^{-i {\bf k} \cdot {\bf x} }\,
\Big[ U({\bf x})-1 \Big] \,.
\ee
Now if we substitute the representation of the Fourier
transform of the pion mean-field (\ref{twopieces}) into
expressions (\ref{H-1-sym-res-mp}, \ref{E-1-sym-res-mp})
we see immediately that the delta function piece in eq.~(\ref{twopieces})
does not contribute to $\widetilde H$, which means that
in (\ref{H-1-sym-res-mp}) we can always replace
$\widetilde U$
by its smooth part $\widetilde U_{{\rm smooth}}$.
On contrary in the expression  (\ref{E-1-sym-res-mp})
for $\widetilde E$
the contribution
of the delta function is nonzero and has the form
(we denote this contribution $\widetilde E_{\pi}$):
\be
\label{E-1-sym-res-mp-pi}
&& \widetilde E^{(3)}_{\pi}(x,\xi ,\Delta ^2)
\, =\, \frac{ F(\Delta^2)}{f_\pi^2}
 \; \int \frac{d^4p}{(2\pi )^4} \;
\delta \left[ \Bigl(x- \xi \Bigr) M_N-v\cdot p \right] \\
\nonumber
&& \, \times \,
\frac{M^{1/2} \Big( ( p+\Delta)^2 \Big)}{ \Big( (p+\Delta)^2-M^2+i0 \Big)} \,
 \,
\frac{M^{3/2} \Big( ( p-\Delta)^2 \Big)}{\Big( (p-\Delta)^2-
M^2+i0 \Big) }
\, +\, \Biggl( \xi \to -\xi, \;\; \mbox{{\boldmath $\Delta$}}
\to -\mbox{{\boldmath $\Delta$}}
\Biggr)\, ,
\ee
where we introduced the following form factor:
\be
F(-\vec k^2) \, =\, \frac{4 M_N^2 f_\pi^2}{3k^3}\int d^3 x\ \exp(i\vec k
\cdot \vec x) \, \, {\rm Tr} \Big[ \Big( U(\vec{x})-1 \Big) \tau^3 \Big]\, .
\label{Fff}
\ee
Now the crucial observation is that the integral over $p$ coincides
$exactly$ (up to trivial renaming of variable) with the expression
for the light-cone pion distribution amplitude in the instanton model
of the QCD vacuum \cite{pp,ppr}\footnote{More precisely, with
distribution amplitude of virtual pion with virtuality $\Delta^2$.}.
Therefore the expression
(\ref{E-1-sym-res-mp-pi}) for the $\widetilde E_\pi$ can be written
in the compact form:
\be
\widetilde E^{(3)}_\pi(x,\xi,\Delta^2) \, =\, \frac{F(\Delta^2)}{|\xi|}\,
\theta \Big(|x|<|\xi| \Big)
\Phi_\pi\Biggl(\frac{x}{\xi},\Delta^2\Biggr),
\label{pionpoler}
\ee
where $\Phi_\pi(z,\Delta^2)$ is the
distribution amplitude of virtual pion, normalized
by
\be
\int_{-1}^1 dz \Phi_\pi(z,\Delta^2)=1.
\ee
Generalizing slightly the technique of refs.~\cite{pp,ppr},
we computed the $\Delta^2$ dependence of the virtual pion
distribution amplitude. At small $\Delta^2$ it has the form:

\be
\Phi_\pi(z,\Delta^2)=
\Biggl(1-\frac{N_c (\Delta^2-m_\pi^2)}{24\pi^2 f_\pi^2}\Biggr )
\Phi_\pi(z) \,+\,  \frac34 (1-z^2)
\frac{N_c (\Delta^2-m_\pi^2)}{24\pi^2 f_\pi^2} \, +\,
\ldots
\label{virpion}
\ee
The pion distribution amplitude
calculated in the instanton model of QCD vacuum
\cite{pp,ppr} is very close to the asymptotic one
$\Phi_\pi(z)=\Phi_\pi(z,\Delta^2\to m_\pi^2)=\frac34 (1-z^2)$.
Therefore from the eq.~(\ref{virpion}) we can conclude that
the dependence of the distribution amplitude
on the virtuality of the pion
is rather weak. In addition, under evolution this dependence
disappears and we have asymptotically $\Phi_\pi(z,\Delta^2)_{asy}=
\frac{3}{4} (1-z^2)$. In what follows we shall therefore
drop the dependence
of distribution amplitude of the virtual pion on $\Delta^2$.

Using the small momentum asymptotic of the pion mean-field
(\ref{Ytail})
one gets immediately the small $\Delta^2$ asymptotic of the form factor
$F(\Delta^2)$ of eq.~(\ref{Fff}):
\be
\lim_{\Delta^2\to m_\pi^2} F(\Delta^2) \, =\,
- \, \frac{4 g_A M_N^2}{(\, \Delta^2-m_\pi^2 \,)}\, .
\label{pionpole}
\ee
The expression (\ref{pionpole})
yields then  the pion pole contribution
to the SPD $\widetilde E^{(3)}$:
\be
\lim_{\Delta^2\to m_\pi^2}
\widetilde E^{(3)}_\pi(x,\xi,\Delta^2) \, =\,
-\, \frac{4 g_A M_N^2}{|\xi|(\, \Delta^2-m_\pi^2 \,)}
\, \theta \Big( |x|<|\xi| \Big)
\Phi_\pi\Biggl( \frac{x}{\xi} \Biggr),
\label{pionpoler1}
\ee
and as a consequence of the sum rule (\ref{srE})
the pion pole contribution to the
pseudoscalar
nucleon form factor $G_P^{T=1}(\Delta^2)$:

\be
\lim_{\Delta^2\to m_\pi^2}G_P^{T=1}(\Delta^2) \, =\,
\lim_{\Delta^2\to m_\pi^2}
\int_{-1}^1 dx \, \widetilde E^{(3)}_\pi(x,\xi,\Delta^2) \, =\,
-\, \frac{4 g_A M_N^2}{(\, \Delta^2-m_\pi^2 \,)}\, .
\ee
 We see that the appearance of the pion pole in $\widetilde E^{(3)}$ is
required by spontaneously broken chiral symmetry.
In order to reproduce it in some model
the latter should respect the chiral Ward identities.
For example, in computation of SPD's in the bag model \cite{Ji3}
the chiral Ward identities are violated and the pion pole contribution
(\ref{pionpole}) is missed.
The chiral quark-soliton
model respects all chiral Ward identities what allows to split
$unambiguously$
SPD $\widetilde E^{(3)}$ into two pieces:
\be
\widetilde E^{(3)}(x,\xi,t)=\widetilde E^{(3)}_\pi(x,\xi,t)+
\widetilde E^{(3)}_{\rm smooth}(x,\xi,t)\, ,
\label{poleplussmooth}
\ee
the result for $\widetilde E^{(3)}_{\pi}$ is given by eq.~(\ref{pionpoler}),
the results for $\widetilde E^{(3)}_{\rm smooth}(x,\xi,t)$ is given
by eq.~(\ref{E-1-sym-res-mp}) with
$\widetilde U$ replaced by
its smooth part $\widetilde U_{{\rm smooth}}$.

Let us note that the form factor
$F(t)$ ($t=\Delta^2$)
in a parametrically wide region $m_\pi^2\ll |t| \ll M_N^2$
contains significant contributions other than the simple pion pole
(\ref{pionpole}). Numerically we found that the form factor
$F(t)$ can be parametrized at $|t|\ll M_N^2$ in the following form:

\be
F(t) \, \approx \,  -\frac{4.4}{t-m_\pi^2}
\, \Biggl( 1+\frac{1.7\ (t-m_\pi^2)}{(1-0.5\ t)^2} \, \Biggr)
\, ,
\ee
where all dimensional quantities are in units of GeV.

It is worth mentioning that although we obtain the pion pole contribution
to the spin-flip SPD $\widetilde E$ (\ref{pionpoler1}) in the model
calculation, actually
the existence of this contribution as such
 follows from general considerations of the
chiral Ward identities \cite{NDCEBAF}.

\section{Numerical results and discussion}
\label{section_numeric}
We have calculated numerically the isovector distributions
$\widetilde H^{(3)}(x,\xi,\Delta^2)$ and
$\widetilde E^{(3)}(x,\xi,\Delta^2)$.
For the calculations we use the variational estimate of
the soliton profile, eq.(\ref{hedge}), of ref.\cite{DPP}
($M_0 = 350\, {\rm MeV}$),
\beq
P(r) \;\; = \;\; -2\;\arctan\left(\frac{r_0^2}{r^2}\right) ,
\hspace{1cm}
r_0 \;\; \approx \;\; 1.0/M_0 ,
\hspace{1cm} M_N \;\; \approx \;\; 1170\;{\rm MeV} ,
\la{varprof}
\eeq
which has been used in the calculation of usual parton distributions
in refs.\cite{DPPPW}. Furthermore, we approximate the
momentum--dependent mass predicted by the instanton model of the
QCD vacuum \cite{DP} by the simple form
\be
M(-p^2) &=& \frac{M_0 \Lambda^{6}}{(\Lambda^2+p^2)^3},
\ee
where the parameter $\Lambda$ is related to the averaged instanton
size, $\rho$, by $\Lambda = 6^{1/3} \rho^{-1} $. This
expression reproduces
the asymptotic behaviour of $M(p^2)$ at large euclidean $p^2$
obtained in the instanton vacuum,
\be
\nonumber
\hspace{3cm} M(-p^2) &\sim& \frac{36 M_0}{\rho^6 p^6}\, , \hspace{2cm}
(p^2 \rightarrow \infty ).
\ee
We have explored also other forms of the momentum dependence of the mass
and found that numerically the results are very close to each other.

\subsection{Results for $\widetilde H (x,\xi,\Delta^2)$}

We estimate the Dirac continuum contribution to
$\widetilde H^{(3)}(x,\xi,\Delta^2)$
using the interpolation formula, eq.~(\ref{H-1-sym-res-mp}), which
gives a reliable approximation preserving all qualitative features
of the continuum contribution. The contribution of the discrete level
is calculated using eq.~(\ref{valence1}).

First we compute $
\widetilde H^{(3)}(x,\xi,\Delta^2)$ in the forward limit, $\Delta\to
0$, where it coincides with the usual quark and antiquark
distributions:
\be
\widetilde H^{(3)} \Big( x, \xi = 0, \Delta^2 = 0 \Big) &=&
\left\{
\begin{array}{cr}
\Delta u(x) -\Delta d(x),& \hspace{.5cm} x \; > \; 0\,, \\
\Delta \bar u(-x)-\Delta \bar d(-x)
,& \hspace{.5cm} x \; < \; 0 \,.
\end{array}
\right.
\label{forward_limit1}
\ee
The result is shown in Fig.~1, where we plot
separately the contributions
of the discrete level and that of Dirac continuum (computed from the
interpolation formula), as well as their sum.
The forward limit reproduces the polarized quark distribution obtained
in \cite{DPPPW} in the same model and at the same level of approximation.
{}From Fig.~1 we see that the Dirac continuum contribution leads to
new qualitative prediction for polarized quark distribution:
the existence of large flavour asymmetry of antiquark distribution
$\Delta \bar u(x) -\Delta \bar d(x)$,
the feature which was noted first in \cite{DPPPW} (see also
\cite{dubna}). To our best knowledge all parametrizations of
polarized quark distributions assume flavour symmetric
antiquark distributions.

The importance of the Dirac continuum contribution to
$\widetilde H^{(3)}(x,\xi,\Delta^2)$ is also shown on Fig.~2. We plot
there separately the discrete level and continuum contributions for
$\Delta^2=-0.5$~GeV$^2$ and $\xi=0.2$. Comparing Fig.~1 and Fig.~2 we
see that the $\xi$-dependence of $\widetilde H^{(3)}(x,\xi,\Delta^2)$
is mostly due to the $\xi$-dependence of
Dirac continuum contribution.
Since the Dirac continuum contribution is symmetric in variable $x$,
we can expect strong $\xi$-dependence
only in part of $\widetilde H^{(3)}$ which is even in $x$ .
This observation allows us to conclude that with good accuracy we can
put (about $\Delta^2$ dependence see below)
$$\widetilde H^{(3)}(x,\xi)-
\widetilde H^{(3)}(-x,\xi) \,  \approx \,
\Delta u(x) -\Delta \bar u(x)
-\Delta d(x) +\Delta \bar d(x)\, .$$
The message which might be useful for modelling of SPD's in terms
of double distributions \cite{Rad,Rad2}.
Additionally we see that ``antiquark" skewed distribution ($\widetilde
H^{(3)}$ at negative $x$) is large and it is dominated by the Dirac
continuum contribution.

In order to illustrate the dependence of
$\widetilde H^{(3)}(x,\xi,\Delta^2)$ on
$\xi$ and $\Delta^2$ we plot this function for a fixed momentum transfer
of $\Delta^2=-0.5$~GeV$^2$ for various values of $\xi$ (see
upper panel of Fig.~3),
and for fixed $\xi = 0.2$ and various values of momentum transfer
(see lower Fig.~3). We clearly see from the figures that the
distribution has ``cusps" at $x=\pm \xi$.

Using the results for $\widetilde H^{(3)}(x,\xi,\Delta^2)$ we computed
$\xi$ and $\Delta^2$ dependence of its Mellin moments. Lorentz invariance
requires that the $N$-th Mellin moment should be a polynomial
of order $N$ in $\xi$ \cite{JiReview}:

\be
\int_{-1}^1dx\ x^{N-1} \widetilde H^{(3)}(x,\xi,\Delta^2) \, =\,
\sum_{k=0}^{[N/2]}\xi^{2k} \, \,  h^{N}_k(\Delta^2).
\ee
Here we prefer to use decomposition of Mellin moments in
partial waves of $q\bar q$ pairs in $t-$channel \cite{MVP98}:

\be
\int_{-1}^1dx\ x^{N -1} \widetilde H^{(3)}(x,\xi,\Delta^2) \, =\,
\xi^N \sum_{l=0}^{N}  P_l \Biggl(\,\frac{1}{\xi}
\, \Biggr) \, a^{(N)}_l(\Delta^2).
\label{pwaofmoments}
\ee
Where $P_l(x)$ are Legendre polynomials and
$l$ is an angular momentum of exchanged $q\bar q$ pair,
it runs over (odd) even values for (odd) even $N$.
If we now take the forward limit in eq.~(\ref{pwaofmoments}) we
obtain:

\be
\int_{-1}^1dx\ x^{N -1} \Big[ \Delta u(x)-\Delta d(x) \Big] \, =\,
\frac{1}{2^N}\ \frac{\Gamma(2  N+1)}{\Gamma(N+1)^2} \, a^{(N)}_N(0).
\ee

Qualitatively we may expect that the slope of
$\Delta^2$ dependence of the form factor
$a^{(N)}_l(\Delta^2)$ is governed by the mass of a low-lying
isovector resonance with spin $l$ and  unnatural parity.
This dependence can be phenomenologically described by simple dipole
fit:

\be
a^{(N)}_l(\Delta^2)=  \frac{a^{(N)}_l(0)}{\Big( \, 1-\Delta^2/M_l^{(N) 2}\,
\Big)^2} \, ,
\label{dipolefit}
\ee
where $M_l^{(N)}$ is a phenomenological parameter (dipole mass).
The results of the calculation and of the fits are given in Table~1.
One should note that $a_1^{(1)}(0)$ is identical to the usual
axial coupling constant $g_A$ and $M_1^{(1)}$ to the corresponding dipole
mass parameter. We see that our theoretical values of these parameters
are very close to experimental ones. One should not overestimate this
in view of the approximations done in the present approach
(interpolation formula, no rotational $1/N_c$ corrections, etc.).
Still the result shows that the method is well founded.
As far as the other $a_l^{(N)}(\Delta^2)$
concerns, we see that the chiral quark-soliton model reproduces the
qualitative expectation that the dipole mass in the form factors
$a_l^{(N)}(\Delta^2)$, describing an exchange with angular momentum
$l$, in fact increasing with $l$.

In modelling of $\Delta^2$-dependence of SPD's usually the
factorization ansatz
$\widetilde H(x,\xi,\Delta^2)=\widetilde H(x,\xi ) G_A(\Delta^2)$
is used. This ansatz would imply e.g. that dipole masses $M_l^{(N)}$
are independent of $N$ and $l$. By explicit calculation in the chiral
quark-soliton model we demonstrated that this is not the case.

Let us note that the hard exclusive reactions can be viewed as
a ``tool" to create fundamental
probes which are absent in the Nature.
For example, among the form factors $a_l^{(N)}(\Delta^2)$
only one ($N=1$, $l=1$) can be measured by a probe provided
by the Nature ($W, Z$ bosons), all others are not accessible
for  electroweak probes. Since in hard exclusive reactions
we can extract the higher spin form factors, this allows us
to study {\it low energy} observables with probes of any spin.
In this respect the hard exclusive reactions can be used not
only for checking of predictions of perturbative QCD but also
as a new way to study low energy properties of hadrons.

\begin{table}
\begin{tabular}{|c|c|c|}
\hline
$(Nl)$ & $a_{l}^{(N)}(0)$ & $M_l^{(N)}$~GeV\\
\hline
(11)& 1.25 & $ 0.9$       \\
(20)& 0.06 & $ 0.9$       \\
(22)& 0.12 & $ 1.1$       \\
(31)& 0.08 & $ 0.9$       \\
(33)& 0.04 & $ 1.4$       \\
\hline
\end{tabular}
\caption{ Values of parameters of dipole fit
for Mellin moments of isovector
$\widetilde H$, see eq.~(\ref{dipolefit})}
\end{table}

\subsection{Results for $\widetilde E (x,\xi,\Delta^2)$}

The skewed distribution $\widetilde E^{(3)}(x,\xi,\Delta^2)$
is dominated by the pion pole contribution (\ref{pionpoler}).
We computed also the smooth part of $\widetilde E^{(3)}(x,\xi,\Delta^2)$
(see eq.~(\ref{poleplussmooth})). The results are presented in
Figure~4, where we plot the pole and the smooth parts separately
at various values of $\xi$ and $\Delta^2$. We see that the pole
contribution dominates
$\widetilde E^{(3)}(x,\xi,\Delta^2)$ in large range of $\Delta^2$
and $\xi$.
The results for the total distribution
(pole+non-pole) at $\Delta^2=-0.5$~GeV$^2$
and various values of $\xi$ are presented on Fig.~5.
The $\xi$ dependence of Mellin moments of
$\widetilde E^{(3)}(x,\xi,\Delta^2)$ has also polynomial form and
we checked that our model calculations reproduce this feature.

The arguments presented here for $\widetilde E$
can be easily extended to the analogous SPD for
$N\to Y$ SPD's ($Y$ is a hyperon from the octet $Y=\Lambda,\Sigma$).
In this case the flavour changing $\widetilde E$ has a contribution
from of the kaon pole of the form:

\be
\widetilde E^{N\to Y} \,=\,
-\frac{4 g_1^{N\to Y} M^2_N}{|\xi|(\, \Delta^2-m_K^2 \,)}
\, \theta \Big( |x|<|\xi| \Big)
\Phi_K\Biggl( \frac{x}{\xi}\Biggr),
\ee
where $\Phi_K$ is kaon distribution amplitude and
$g_1^{N\to Y}$ the constant entering in description of
the semileptonic decays of hyperons.

\subsection{Comparison with other models}

Helicity skewed quark distributions were computed previously
in the bag model \cite{Ji3}. Unfortunately in this model the chiral
symmetry is broken explicitly by boundary conditions at bag surface.
Therefore the crucial contribution of the pion pole to
$\widetilde E^{(3)}(x,\xi,\Delta^2)$ is missed in this model.
Nevertheless the bag model describes
qualitative features of the skewed quark
distributions for
which the ``resonance part"
\footnote{The notion of virtual hadron
can not be defined in QCD apart from special cases (large $N_c$
limit, pions, etc.).
We use the term ``resonance part" to denote specific contributions
to SPD arising only in non-forward limit.}
is relatively small, these are
odd (even) in $x$ part of isovector (isoscalar)
$\widetilde H(x,\xi,\Delta)$ and $\widetilde E(x,\xi,\Delta)$.

In another approach,
proposed by Radyushkin \cite{Rad,Rad2}, one writes a spectral
representation for the matrix element of the light--ray operator
in terms of a so-called double distribution.
The skewed distribution for a given value of $\xi$
is then obtained as a particular one--dimensional reduction
of this two--variable distribution.
The advantage of this approach is
that the resulting skewed parton distribution satisfy automatically
polynomiality conditions (\ref{pwaofmoments}). However, as it was shown
in ref.~\cite{PW99} the parametrization of skewed quark distributions
in terms of double distributions is not complete.
This incompleteness can be seen especially clear if one considers the
Mellin moments of skewed quark distributions. For example, the expression
for $\widetilde H(x,\xi,\Delta)$ in terms of double distributions
requires that its $N$-th Mellin moment is a polynomial of order
$N-2$ in variable $\xi$, what is
in contradiction with (\ref{pwaofmoments}) for
even $N$. This problem can be easily cured
if one adds one additional function to the double distribution
parametrization of light-cone nucleon matrix elements, see \cite{PW99}.

As we saw in our model the even moments of  $\widetilde H(x,\xi,\Delta^2)$
are almost $\xi$-independent because the contribution of the
Dirac sea drops out. Owing to this feature of
$\widetilde H(x,\xi,\Delta^2)$ the
additional function which one needs to add to the double distribution
parametrization of light-cone nucleon matrix elements is very small.

\section{Conclusions}
We have shown that the helicity skewed distribution $\widetilde E$
is dominated in large range of $\Delta^2$ and $\xi$ by contribution
of the pion pole (\ref{pionpoler1}). This result can be viewed as
generalization of well known chiral Ward identities for local
currents to bilocal quark operators on the light cone.
The fact that $\widetilde E$ is fixed to great extent
by the pion pole contribution opens the possibility to measure
$\widetilde H$ by choosing observables which are proportional
to the product $\widetilde H\cdot \widetilde E$. One of examples of such
quantity is azimuthal spin asymmetry in hard exclusive
production of pions and kaons \cite{FPPS}.

The skewed quark distribution $\widetilde H$ has been computed in wide
range of $\Delta^2$ and $\xi$ using chiral quark-soliton model.
We have demonstrated that contribution of the Dirac continuum
is crucial to describe transition between two
regions $|x|>\xi$ and
$|x|<\xi$. Also we saw that the $\xi$ dependence of the SPD's
is mostly due to Dirac continuum contribution. Since the Dirac
continuum contribution to $\widetilde H^{(3)}$ is symmetric in variable
$x$, we can expect that the dependence of the combination
$\widetilde H^{(3)}(x,\xi)-\widetilde H^{(3)}(-x,\xi)$ on $\xi$
is rather weak.
This, in particular, implies that in modelling
of SPD's in terms of double distributions the function
$h(x,y)$\footnote{See for notations \cite{Rad,Rad2}.} should
strongly depend on $C$ parity of the SPD.

Studying the $\Delta^2$ dependence of the helicity skewed parton
distributions we have seen that the factorization ansatz
$\widetilde H(x,\xi,\Delta^2)=\widetilde H(x,\xi) G_A(\Delta^2)$
is in contradiction with our calculations.

For models which use the usual quark distributions to model
SPD's (see e.g. \cite{Rad,Rad2,Andr}) one can use
replacement of the slope of Regge trajectory
in small $x$ parametrization of $\Delta q(x)\sim 1/x^{\alpha_0}$ at
low normalization point by Regge trajectory $\alpha_0\to
\alpha_0+\alpha'\ \Delta^2$. Such replacement describes
qualitatively correct the
$\Delta^2$ dependence of the Mellin moments of SPD's as was discussed
in present paper.

\vspace{1cm}
\noindent
{\large\bf Acknowledgements}
\\[.3cm]
We are grateful to Ingo B\"ornig for discussions and help with
numerical calculations.
We are also grateful to L.~Frankfurt, L.~Mankiewicz, G.~Piller,
P.V.~Pobylitsa, A.~Radyushkin, A.~Sch\"afer,  M.~Strikman,
M.~Vanderhaeghen and C.~Weiss for inspiring conversations.
This work has been supported in parts by Deutsche
Forschungsgemeinschaft (Bonn), BMFB (Bonn) and by COSY (J\"ulich).

\newpage
\appendix
\renewcommand{\theequation}{\Alph{section}.\arabic{equation}}
\section{Bound-state level contribution to
$\widetilde H^{(3)}(x,\xi,\Delta^2)$ and
$\widetilde E^{(3)}(x,\xi,\Delta^2)$}
\setcounter{equation}{0}

We present here the contributions of the discrete bound-state
level to the isovector $\widetilde H^{(3)}(x,\xi,\Delta^2)$
and $\widetilde E^{(3)}(x,\xi,\Delta^2)$.
The bound-state level occurs in the grand spin $K=0$ and parity
$\Pi=+$ sector of the Dirac Hamiltonian (\ref{hU}). In that sector
the eigenvalue equation takes the form:

\be
\left(\begin{array}{cc}
 M \cos P(r) &
{\displaystyle -\frac{\partial}{\partial r}-\frac{2}{r} - M \sin P(r)}\\
{\displaystyle \frac{\partial}{\partial r} - M \sin P(r)} & - M \cos P(r)
\end{array}\right)
\left(\begin{array}{c}
h_0(r) \rule[-.5em]{0cm}{2em} \\ j_1(r) \rule[-.5em]{0cm}{2em}
\end{array}\right)
&=& E_{\rm lev}
\left(\begin{array}{c}
h_0(r) \rule[-.5em]{0cm}{2em} \\ j_1(r) \rule[-.5em]{0cm}{2em}
\end{array}\right).
\label{H-K-0:Pi-plus}
\ee
We assume that the radial wave functions are normalized by the
condition

\be
\int\limits_{0}^{\infty} dr \, r^2  \, \Big[\, h_0^2(r) + j_1^2(r)\, \Big] &=& 1.
\ee
We introduce the Fourier transforms of the radial wave functions,

\be
h(k) &=& \int\limits_{0}^{\infty} dr \, r^2\, h_0(r) R_{k0}(r),
\hspace{1.2cm}
j(k) \;\; = \;\; \int\limits_{0}^{\infty} dr \, r^2\, j_1(r) R_{k1}(r),
\ee
where
\be
R_{kl}(r) &=&  \sqrt{ \frac{k}{r}} J_{l+\frac12}(kr)
\;\; = \;\; (-1)^l\sqrt{\frac{2}{\pi}} \frac{r^l}{k^l}
\left( \frac{1 }{r } \frac{d}{dr } \right)^l \frac{\sin kr }{r}.
\ee

The bound-state level contribution to the  $\widetilde H$ and
$\widetilde E$
distribution function can be simply obtained from the general
eqs.~(\ref{H-singlet-general}, \ref{E-nonsinglet-general}).
Compactly the corresponding expressions can be written as:

\be
\nonumber
\widetilde H^{(3)}_{\rm lev}\, (x,\xi ,\Delta ^2)\, \delta^{3b} &-&
\widetilde E^{(3)}_{\rm lev}\, (x,\xi ,\Delta ^2)
\cdot \frac{\Delta^3\Delta^b}{\Delta^2} \,=\, \frac{2 N_c M_N\pi}{3}
\int \frac{d^2{\bf k_\perp}}{(2\pi)^2} \, \frac{1}{k k'} \\
\nonumber
\times \Biggl\{ h(k)h(k')\delta^{3b}
\label{valence1}
&-& \Bigl[ h(k)j(k') n^b + h(k' )j(k) n_*^b \Bigr]
+j(k')j(k) \Bigl[n^3n_*^b+n_*^3 n^b-\delta^{3b}(n_*\cdot n)\Bigr] \Biggr\},\\
&&
\ee
where
\be
k&=&\sqrt{{\bf k_\perp}^2+ \Bigl( (x+ \xi)M_N-E_{\rm lev} \Bigr)^2}\\
k'&=&\sqrt{ \Bigl( {\bf k_\perp+
\mbox{{\boldmath $\Delta_T$}}} \Bigr)^2+
\Bigl( (x- \xi)M_N-E_{\rm lev}\Bigr)^2}\, .
\ee
Unit vectors $n$ and $n_*$ have components:

\be
\nonumber
n&=&\frac{1}{k} \, \Big( {\bf k_\perp},(x+ \xi)M_N-E_{\rm lev} \Big) \\
\nonumber
n_*&=&\frac{1}{k'} \,  \Big( {\bf k_\perp+
\mbox{{\boldmath $\Delta_T$}}},\, (x- \xi)M_N-E_{\rm lev} \Big).
\ee
The individual expressions for $\widetilde H$ and $\widetilde E$
can be obtained from eq.~(\ref{valence1}) contracting
index $b$ with
$\left( \delta^{b3}-\frac{\Delta^b\Delta^3}{\Delta^2} \right) $ and
$\Delta_\perp^b$.

\newpage
\begin{figure}
 \vspace{-1cm}
\epsfxsize=16cm
\epsfysize=15cm
\centerline{\epsffile{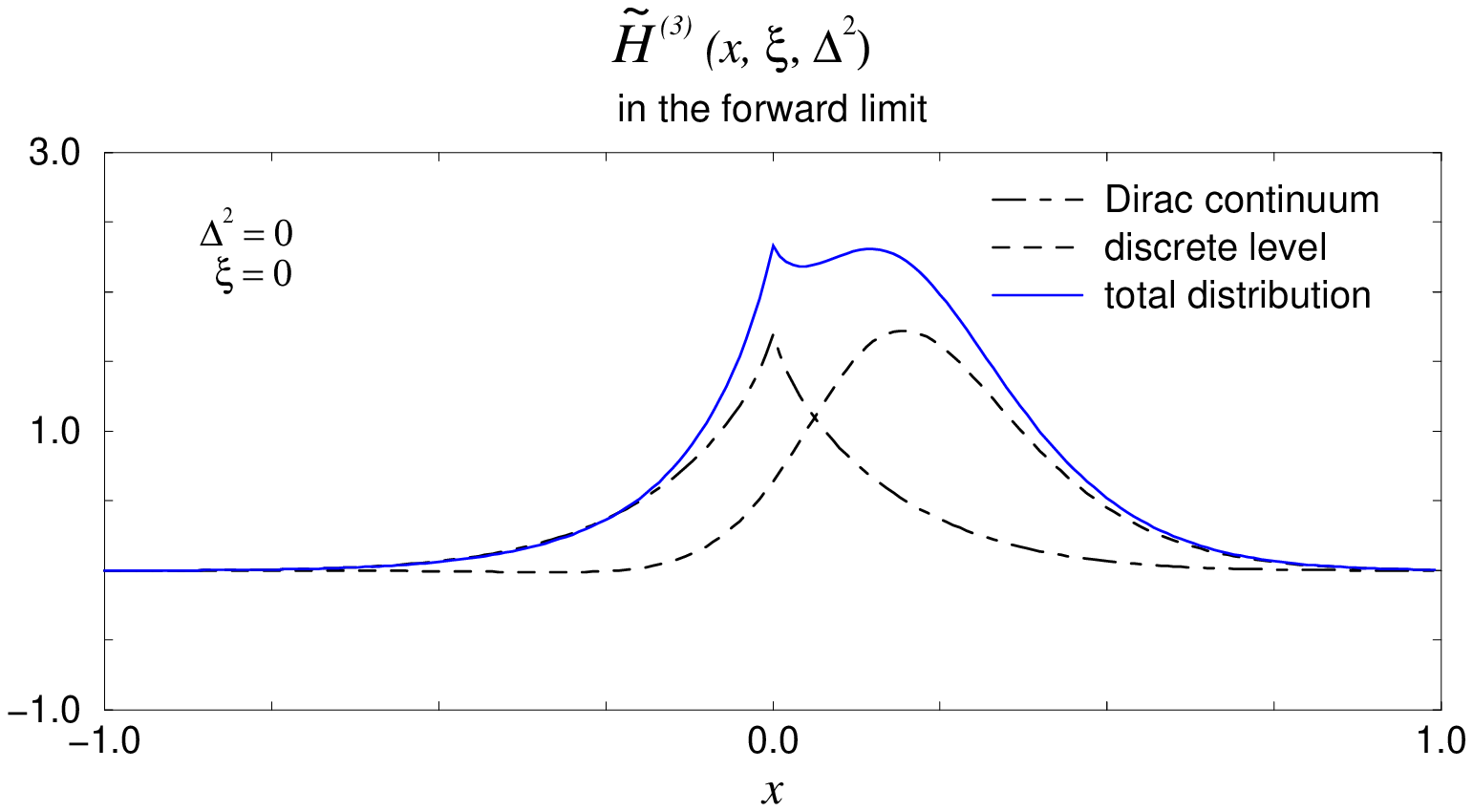}}
\caption[]{
The isovector distribution
$\widetilde H(x,\xi,\Delta^2)$ in the forward
limit, $\Delta=0$.
{\em Dashed line}:
contribution from the discrete level. {\em Dashed-dotted line}:
contribution from the Dirac continuum according to the interpolation
formula, eq.~(\protect\ref{H-1-sym-res-mp}).
{\em Solid line}: total distribution (sum of the dashed
and dashed-dotted curves).}
\end{figure}

\newpage
\begin{figure}
 \vspace{-1cm}
\epsfxsize=16cm
\epsfysize=15cm
\centerline{\epsffile{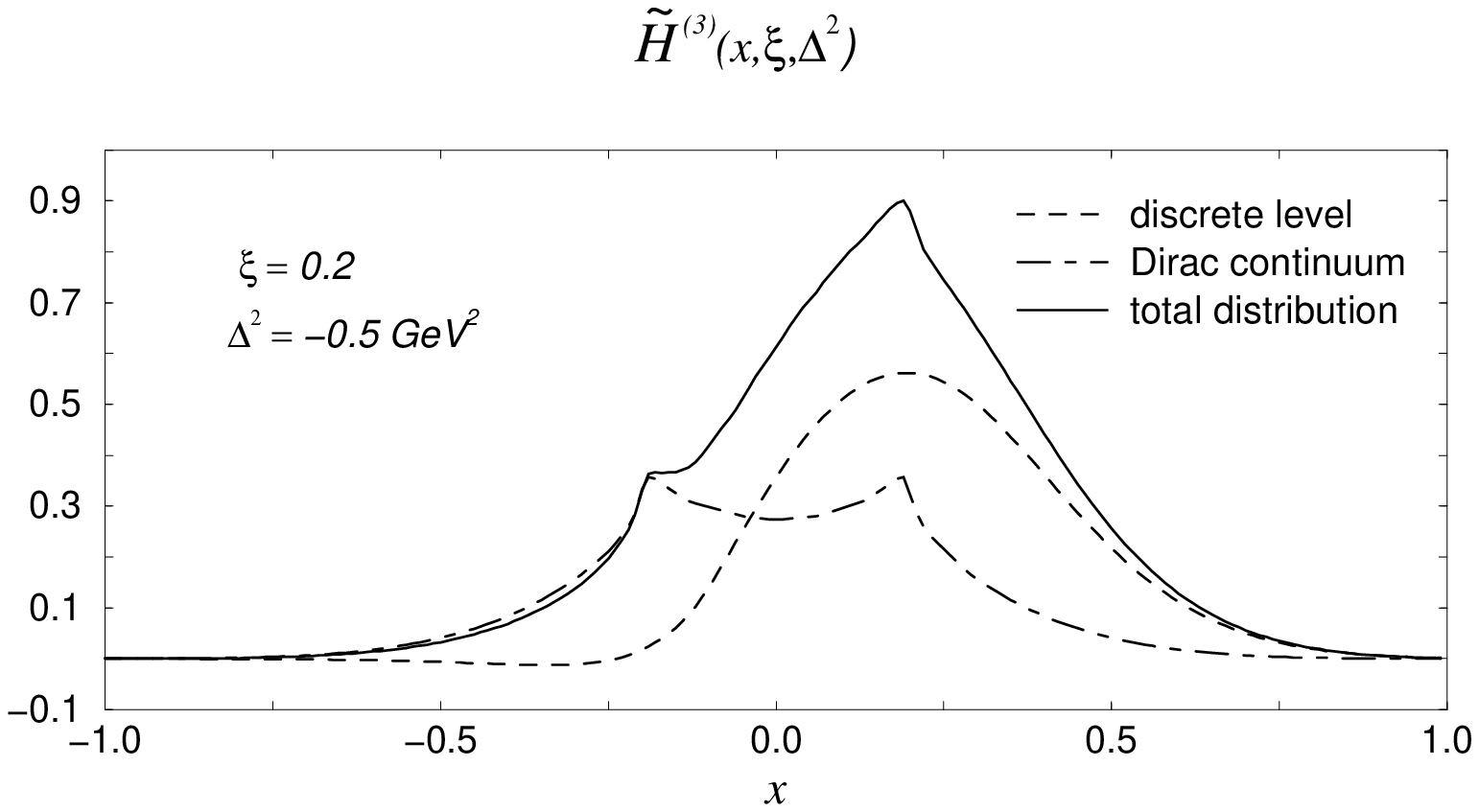}}
\caption[]{
The same as Fig.~1 but for no-forward case $\Delta^2=-0.5$~GeV$^2$
and $\xi=0.2$.
} \end{figure}

\newpage
\begin{figure}
 \vspace{-1cm}
\epsfxsize=16cm
\epsfysize=15cm
\centerline{\epsffile{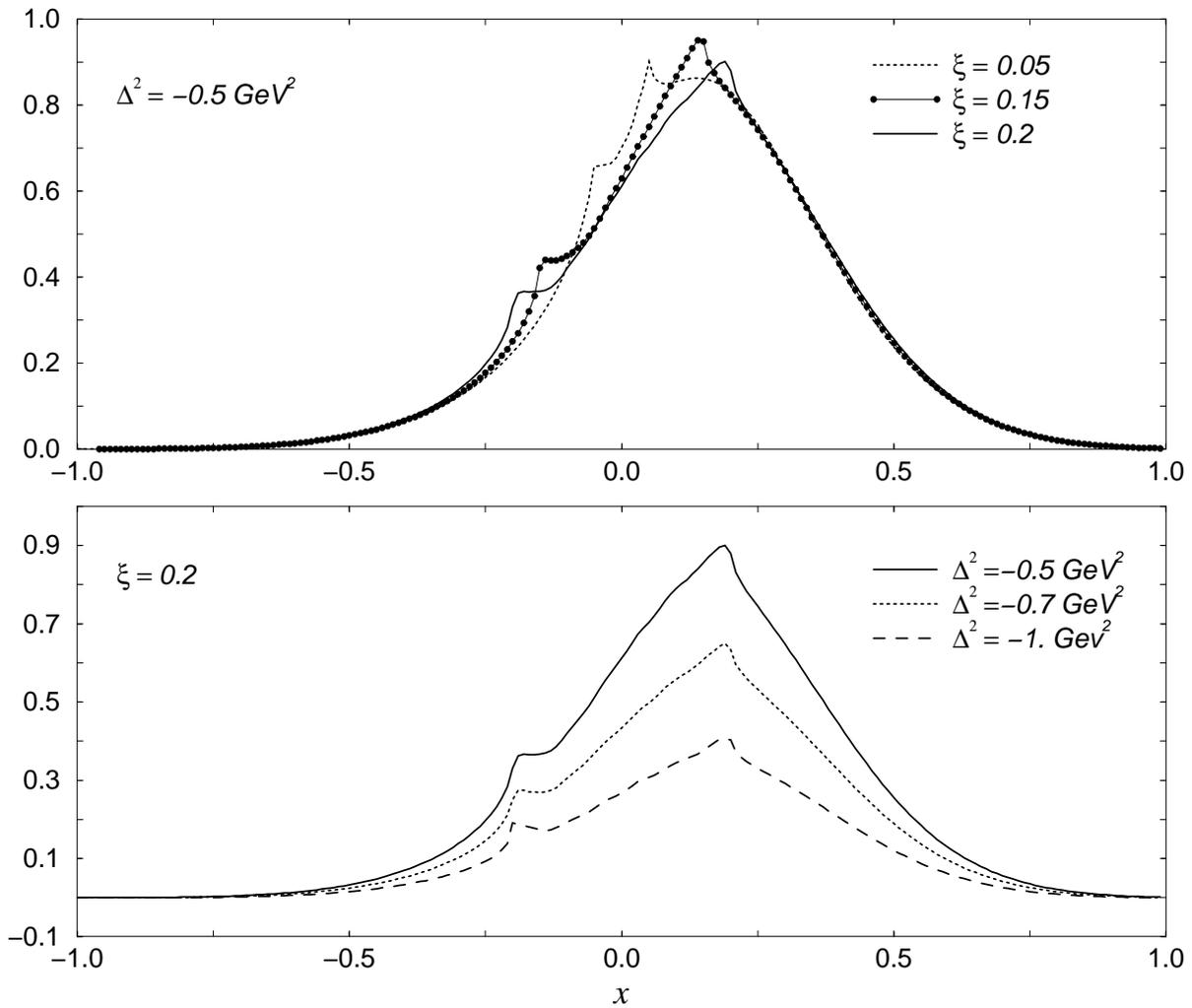}}
\caption[]{ Calculated isovector SPD $\widetilde H$ at various values
of $\Delta^2$ and $\xi$.
} \end{figure}

\newpage
\begin{figure}
 \vspace{-1cm}
\epsfxsize=16cm
\epsfysize=15cm
\centerline{\epsffile{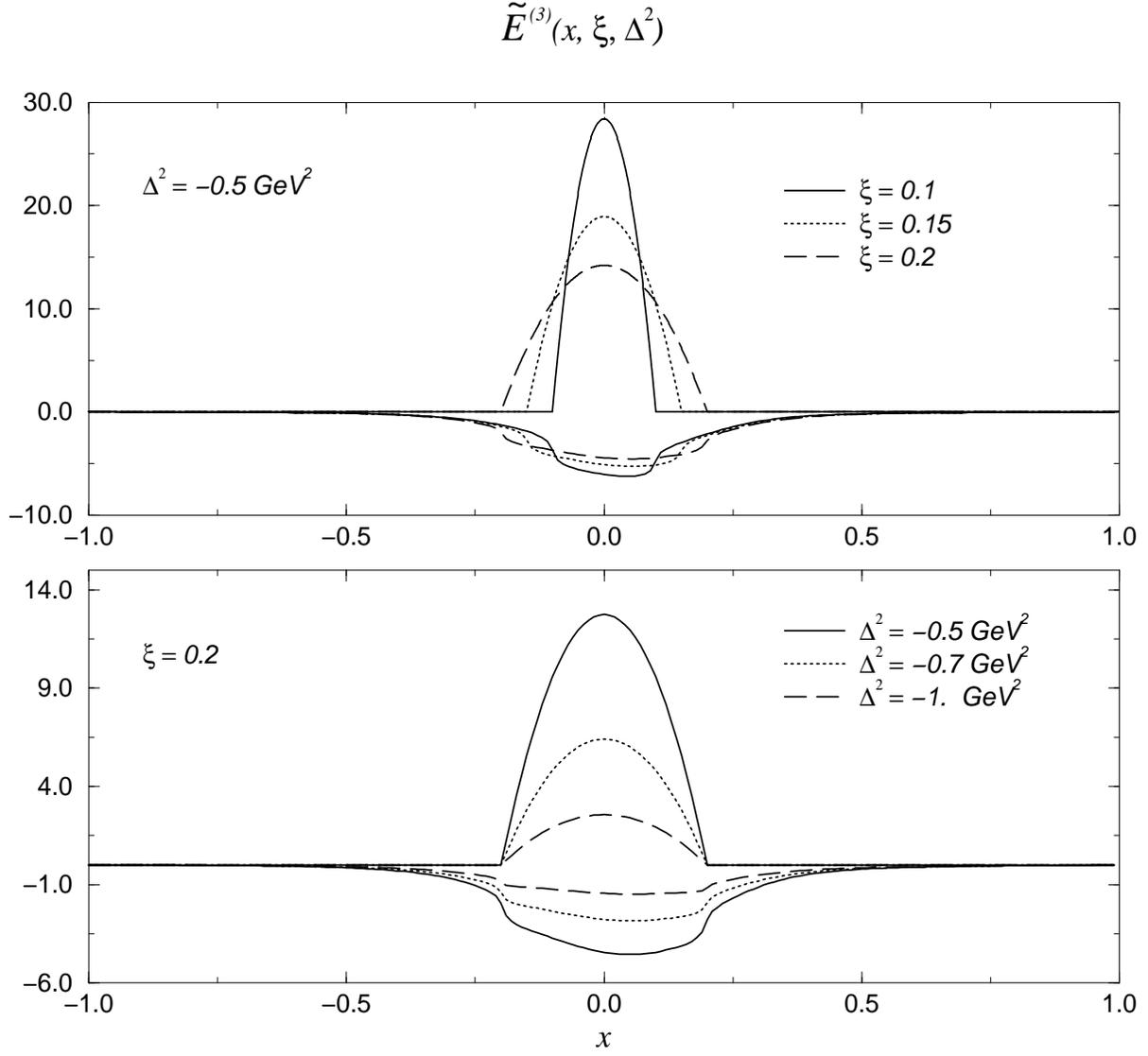}}
\caption[]{
Comparison of pion pole contribution and non-pole part
of isovector $\widetilde E$ at various values of $\Delta^2$ and $\xi$.
The positive curves correspond to pion pole contributions.
} \end{figure}

\newpage
\begin{figure}
 \vspace{-1cm}
\epsfxsize=16cm
\epsfysize=15cm
\centerline{\epsffile{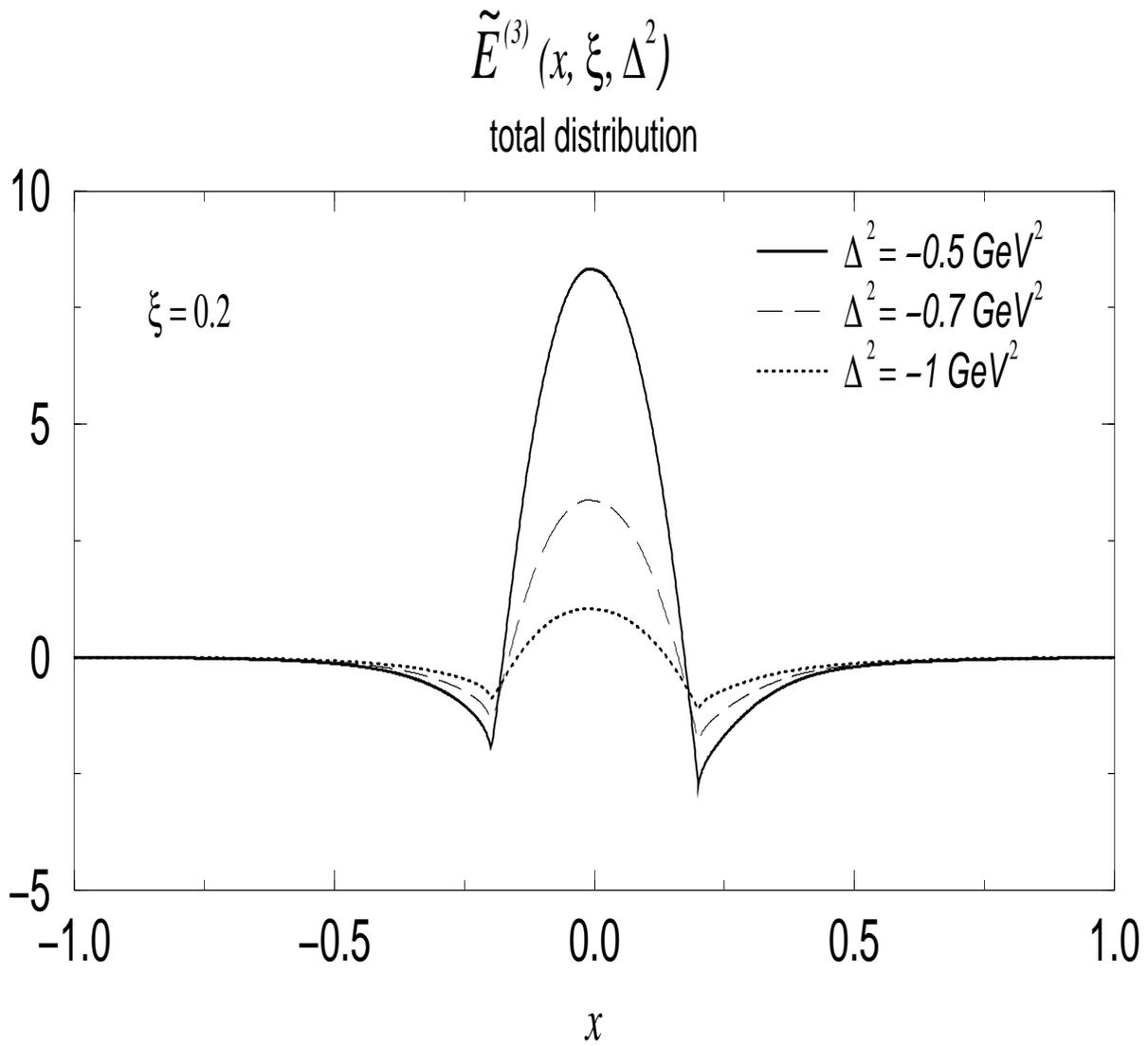}}
\caption[]{
Total result (pole+non-pole) for isovector $\widetilde E$
at $\xi=0.2$ and various values of $\Delta^2$.
} \end{figure}
\end{document}